# Xcos on Web як перспективний засіб навчання моделювання технічних об'єктів бакалаврів електромеханіки

Євгеній Олександрович Модло
Кафедра автоматизованого управління металургійними процесами та електроприводом, ДВНЗ «Криворізький національний університет», вул. Революційна, 5, м. Кривий Ріг, 50006, Україна
eugenemodlo@gmail.com

Сергій Олексійович Семеріков
Кафедра інженерної педагогіки та мовної підготовки,
ДВНЗ «Криворізький національний університет»,
вул. XXII Партз'їзду, 11, м. Кривий Ріг, 50027, Україна
semerikov@gmail.com

**Анотація.** *Цілі дослідження*: визначення перспективного засобу навчання імітаційного моделювання майбутніх бакалаврів електромеханіки.

*Завдання дослідження*: обґрунтувати доцільність використання системи моделювання Xcos on Web як засобу формування у майбутніх бакалаврів електромеханіки компетентностей з моделювання технічних об'єктів.

*Об'єкт дослідження*: використання систем імітаційного моделювання у навчанні бакалаврів електромеханіки.

*Предмет дослідження*: використання Xcos on Web у навчанні моделювання технічних об'єктів бакалаврів електромеханіки.

Використані *методи дослідження*: аналіз досвіду використання існуючих програмних продуктів.

*Результати дослідження*. Система імітаційного моделювання Xcos on Web є перспективним засобом навчання бакалаврів електромеханіки моделювання технічних об'єктів у хмаро орієнтованому середовищі.

*Основні висновки і рекомендації*:

1. Використання систем імітаційного моделювання, таких як Scilab Xcos, є необхідною складовою професійної підготовки бакалаврів електромеханіки.

2. Хмаро орієнтоване середовище навчання на основі комплексного використання мобільних Інтернет-пристроїв майбутніми електромеханіками сприяє формуванню їх професійних компетентностей.

3. Реалізація у Xcos on Web повної функціональності Scilab Xcos створює умови для переходу у навчанні бакалаврів електромеханіки моделювання технічних об'єктів до використання мобільних Інтернет-пристроїв.

**Ключові слова**: система імітаційного моделювання Scilab Xcos; пакет Scilab; засіб Xcos on Web; мобільні Інтернет-пристрої; навчання бакалаврів електромеханіки моделювання технічних об'єктів.

**Y. O. Modlo[*], S. O. Semerikov[ѣ]. Xcos on Web as a promising learning tool for Bachelor's of Electromechanics modeling of technical objects**

**Abstract**. *Research goals*: to identify the perspective learning simulation tool for Bachelors of Electromechanics.

*Research objectives*: to prove the feasibility of using the simulation system Xcos on Web as a tool of forming of future Bachelors of Electromechanics competence in modeling of technical objects.

*Research object*: the use of imitative simulation systems to learning the Bachelors of Electromechanics.

*Research subject*: the use Xcos on Web in learning modeling of technical objects the Bachelors of Electromechanics.

*Research methods* used: the analysis of existing software usage experience.

*Research results*. The imitative simulation system Xcos on Web is a promising cloud-based learning tool for Bachelor's of Electromechanics modeling of technical objects.

*The main conclusions and recommendations*:

1. The use of simulation systems, such as Scilab Xcos, is a necessary part of Bachelor of Electromechanics professional training.

2. Cloud-based learning environment built on the integrative usage of mobile Internet devices promotes the forming of Bachelor's of Electromechanics professional competencies.

3. Implementation the full Scilab Xcos functionality at Xcos on Web creates conditions for transition in Bachelor's of Electromechanics learning the simulation of technical objects to the use of mobile Internet devices.

**Keywords**: imitative simulation system Scilab Xcos; Scilab package; Xcos on Web tool; mobile Internet devices; learning the Bachelor's of Electromechanics to modeling of technical objects.

**Affiliation:** Department of Automated control of metallurgical processes and electric drives, SIHE «Kryvyi Rih National University», 5, Revoliutsiina str., Kryvyi Rih, 50006, Ukraine[*];

Department of engineering pedagogy and language training, SIHE «Kryvyi Rih National University», 11, XXII Partz'yizdu str., Kryvyi Rih, 50027, Ukraine[*].

E-mail: eugenemodlo@gmail.com[*], semerikov@gmail.com[ѣ].

Одним із традиційних для навчання моделювання засобів ІКТ є системи комп'ютерної математики. Так, у [2] автори вказують на доцільність застосування їх разом із системами підтримки навчання бакалаврів електромеханіки моделювання технічних об'єктів, такими як система Moodle, доповнена розробленим фільтром SageCell [3]. Вказаний фільтр здатен реалізовувати чисельне розв'язування систем диференціальних рівнянь, що описують математичні моделі, безпосередньо у системі підтримки навчання.

Однак використання тільки систем комп'ютерної математики (навіть таких потужних, як SageMath [4]) для навчання моделювання технічних об'єктів бакалаврів електромеханіки є недостатнім, оскільки при синтезі та обчисленні моделей систем керування, елементів електроприводу та ін. використовуються насамперед засоби візуального моделювання, що надають можливість будувати динамічні моделі (дискретні, неперервні та моделі систем із розривами). Це визначає необхідність та доцільність об'єднання традиційних систем комп'ютерної математики зі спеціалізованими бібліотеками для моделювання технічних об'єктів у оболонки для візуального конструювання моделей. При цьому вибір середовища для моделювання повинен урахувати специфіку майбутньої професійної діяльності, якою для бакалаврів електромеханіки є синтез відповідних технічних об'єктів – електромеханічних систем.

Опанування моделювання технічних об'єктів забезпечує теоретичне та практичне наповнення фундаментальної, загально та спеціалізовано-професійної підготовки бакалавра електромеханіки. У зв'язку з цим бажано, щоб середовище для їх моделювання надавало користувачеві доступ не лише до традиційних бібліотек моделювання неперервних та дискретних динамічних систем, а й до бібліотек для електричних машин та силових перетворювачів. Крім того, для досягнення цілі мобільності навчання середовище моделювання повинно мати високий рівень кросплатформенності (зокрема, доступ через Web-інтерфейс) та бути вільно поширюваним.

З метою обґрунтованого вибору середовища моделювання технічних об'єктів для бакалаврів електромеханіки було проведено експертне оцінювання найбільш поширених систем візуального моделювання, результати якого подано у таблиці 1.

На поточний момент найбільшу експертну оцінку має середовище Scilab, переваги використання якого у 2011 році визнало Міністерство національної освіти, вищої освіти і науки Франції, надавши Scilab знак визнання його педагогічної значущості для навчання математики

«Reconnu d'Intérêt Pédagogique» [5].

*Таблиця 1*

**Порівняння середовищ моделювання технічних об'єктів**

| Середовище моделювання | Наявність вільно поширюваної версії («так» – 3 бали, «ні» – 0 балів, «так (з обмеженнями)» – 1 бал) | Кількість підтримуваних операційних систем (за кожну систему – 0,5 бали, за «∞» – 1 бал) | Наявність Web-інтерфейсу («так» – 3 бали) | Наявність бібліотек для моделювання неперервних систем («так» – 5 балів) | Наявність бібліотек для моделювання дискретних систем («так» – 5 балів) | Наявність бібліотек електричних машин («так» – 2 бали) | Наявність бібліотек силових перетворювачів («так» – 2 бали) | Загальна оцінка |
|---|---|---|---|---|---|---|---|---|
| 1 | 2 | 3 | 4 | 5 | 6 | 7 | 8 | 9 |
| Analytica | так (з обмеженнями) | 1 (W*) | так | ні | ні | ні | ні | 4,5 |
| AnyLogic | так | 3 (WML) | ні | ні | ні | ні | ні | 4,5 |
| GoldSim | ні | 1 (W) | ні | ні | ні | ні | ні | 0,5 |
| Insight Maker | так | ∞ (JS) | так | ні | ні | ні | ні | 7 |
| MapleSim | ні | 3 (WML) | ні | ні | ні | так | так | 5,5 |
| Minsky | так | 3 (WML) | ні | ні | ні | ні | ні | 4,5 |
| Rand Model Designer | ні | 1 (W) | ні | так | так | ні | ні | 10,5 |
| Scilab Xcos | так | 3 (WML) | так | так | так | ні | ні | 17,5 |
| Simantics System Dynamics | так | 1 (W) | ні | ні | ні | ні | ні | 3,5 |
| Simile | так (з обмеженнями) | 3 (WML) | так | ні | ні | ні | ні | 5,5 |
| Simulink | ні | 3 (WML) | ні | так | так | так | так | 15,5 |
| Temporal Reasoning Universal Elaboration | ні | 1 (W) | ні | ні | ні | ні | ні | 0,5 |
| Vensim | так | 2 (WM) | ні | ні | ні | ні | ні | 4 |
| VisSim | ні | 1 (W) | ні | ні | ні | ні | ні | 0,5 |
| Wolfram SystemModeler | ні | 3 (WML) | ні | ні | ні | так | так | 5,5 |

* W – Windows, M – macOS, L – Linux, JS – JavaScript

Xcos є доповненням до Scilab, що надає можливості синтезу математичних моделей в галузі механіки, гідравліки, електроніки та електромеханіки. Дане середовище візуального моделювання призначене для розв'язання задач динамічного моделювання систем, процесів, пристроїв, а також тестування та аналізу цих систем. При цьому об'єкт, що моделюється (система, пристрій, процес), подається графічно блок-схемою, що включає блоки елементів системи і зв'язки між ними (рис. 1).

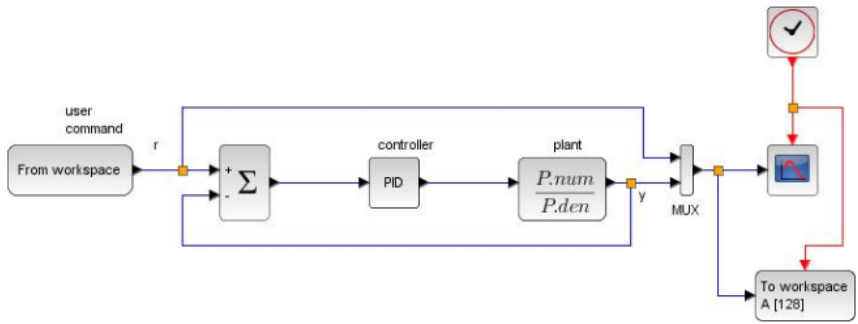

Рис. 1. Приклад моделі у Scilab Xcos

Найбільш вагомими критеріями, що зумовили вибір Scilab як засобу навчання моделювання технічних об'єктів бакалаврів електромеханіки, є наявність бібліотек для моделювання неперервних систем (5 балів), наявність бібліотек для моделювання дискретних систем (5 балів) та наявність Web-інтерфейсу (3 бали). Останнє надає можливість його використання на мобільних Інтернет-пристроях [1].

Сервіс Scilab Cloud (http://scilab.io/services/development/web-application/) надає можливість створювати та оприлюднювати інтерактивні документи, програмна частина яких виконується на боці сервера, що економить ресурси мобільного Інтернет-пристрою. Доступ до Scilab Cloud забезпечується за допомогою будь-якого Інтернет браузера (рис. 2).

За функціональністю даний сервіс є подібним до Wolfram Demonstrations Project, проте, на відміну від останнього, оприлюднювати демонстрації можуть усі користувачі, а не лише адміністратори сервісу, що відповідає моделі поширення документів у SageMathCloud. Проте Scilab Cloud не надає доступ до модуля Xcos.

Сервіс Xcos on web розв'язує цю проблему, надаючи можливість побудови імітаційних моделей технічних об'єктів (зокрема, електромеханічних систем) у мобільному Web-браузері. Поточна мета проекту Xcos on web – відтворити повнофункціональну версію Scilab

Xcos з доступом через мобільний Web-браузер [6].

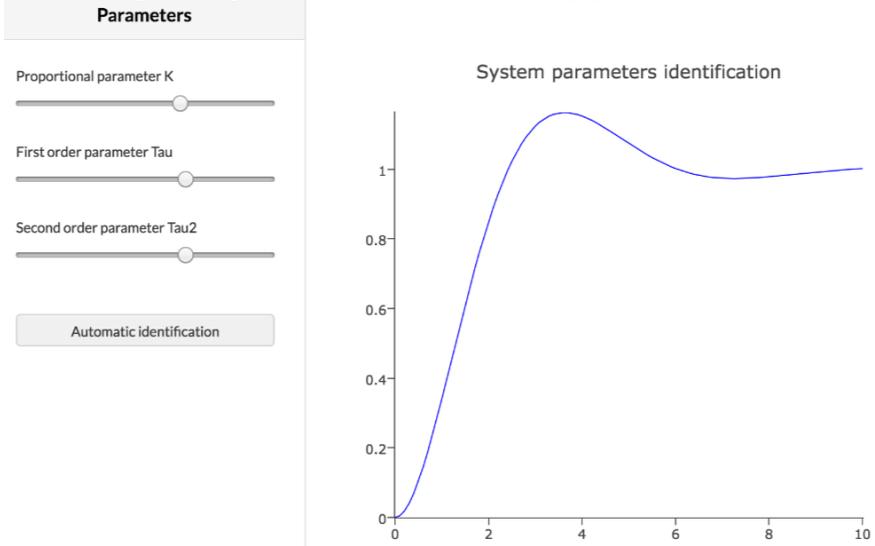

Рис. 2. Приклад інтерактивного документа у Scilab Cloud

Основні компоненти головного вікна Xcos on web подано на рис. 3.

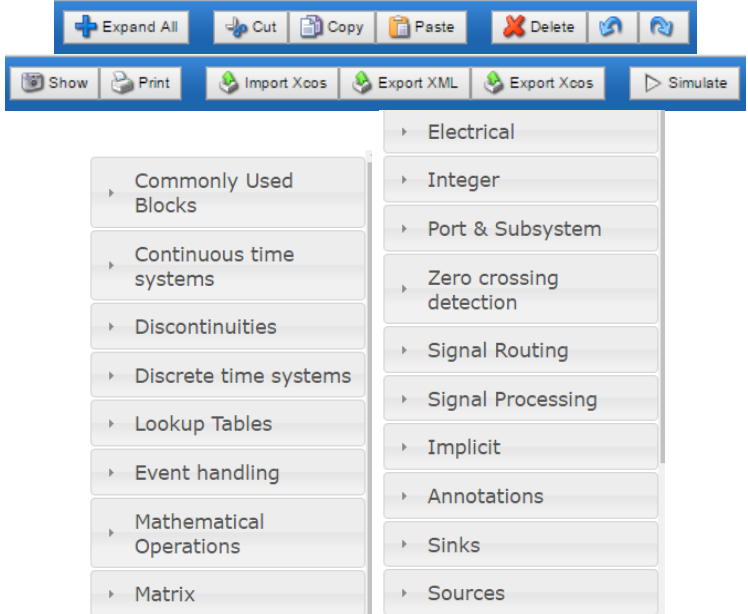

Рис. 3. Компоненти головного вікна Xcos on web

Основну частину головного займає область побудови моделей. Із лівого боку розташована так звана палітра блоків – бібліотека елементів, з яких будується модель. Для використання будь-якого блоку достатньо перетягнути його з палітри у область побудови моделей. Блоки з'єднуються між собою лініями зв'язку.

На рис. 4 показано модель двигуна постійного струму, побудована у Xcos on web.

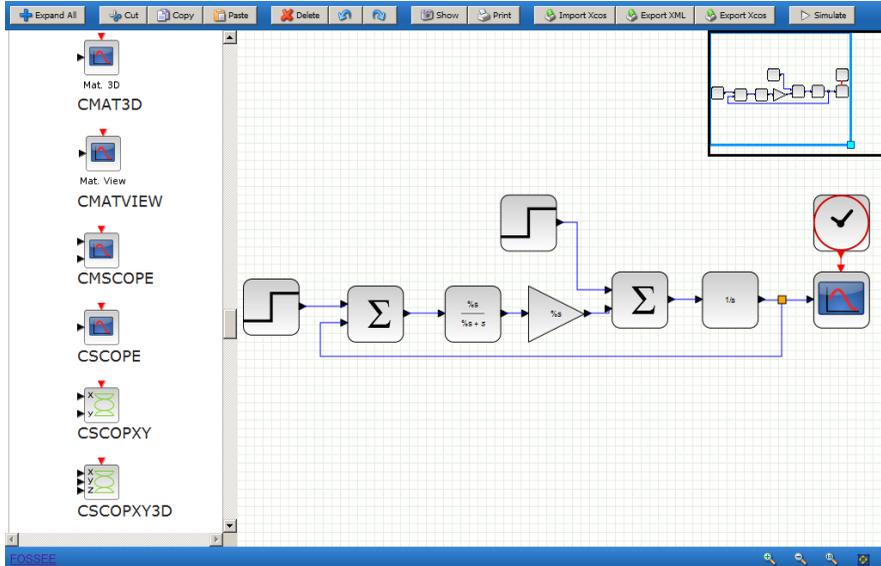

Рис. 4. Модель двигуна постійного струму у Xcos on web

Блоки моделі мають різні параметри, що налаштовуються користувачем подвійним натисканням на обраному блоці. На жаль, проект все ще знаходиться на стадії попередньої розробки, тому не всі налаштування є доступними: так, блоки аперіодичної ланки першого порядку (CLR) та підсилювача (GAIN) у побудованій моделі є неналаштованими, про що свідчить специфікатор формату '%s'. Це унеможливлює проведення експериментів на моделі натисканням кнопки *Simulate*. Тимчасовий варіант обходу цієї проблеми полягає в можливості Xcos on web обміну даними із традиційною версією Scilab Xcos через засоби експорту даних. Фрагмент XML-подання побудованої моделі двигуна постійного струму:

```
<?xml version="1.0" encoding="UTF-8"?>
<XcosDiagram background="-1" title="MavXcos">
    <mxGraphModel as="model">
        <root>
```

```
			<mxCell id="0"/>
			<mxCell id="1" parent="0"/>
			<BasicBlock blockType="c" id="2" interfaceFunctionName="STEP_FUNCTION" parent="1" simulationFunctionName="csuper" simulationFunctionType="DEFAULT" style="STEP_FUNCTION">
...
		</root>
	</mxGraphModel>
  <mxCell id="1" parent="0" as="defaultParent"/>
</XcosDiagram>
```

**Висновки:**

1. Використання систем імітаційного моделювання, таких як Scilab Xcos, є необхідною складовою професійної підготовки бакалаврів електромеханіки.

2. Критеріями вибору середовища моделювання технічних об'єктів є наявність вільно поширюваної версії, кількість підтримуваних операційних систем, наявність Web-інтерфейсу, наявність бібліотек для моделювання неперервних систем, наявність бібліотек для моделювання дискретних систем, наявність бібліотек електричних машин та наявність бібліотек силових перетворювачів.

3. Хмаро орієнтоване середовище навчання на основі комплексного використання мобільних Інтернет-пристроїв майбутніми електромеханіками сприяє формуванню їх професійних компетентностей.

4. Реалізація у Xcos on Web повної функціональності Scilab Xcos створює умови для переходу у навчанні бакалаврів електромеханіки моделювання технічних об'єктів до використання мобільних Інтернет-пристроїв.

**Список використаних джерел**